\documentclass[conference]{IEEEtran}
\ifCLASSINFOpdf
  \usepackage[pdftex]{graphicx}
\else
\fi
\usepackage{flushend}
\usepackage{amsmath}

\begin{document}
%
% paper title
% Titles are generally capitalized except for words such as a, an, and, as,
% at, but, by, for, in, nor, of, on, or, the, to and up, which are usually
% not capitalized unless they are the first or last word of the title.
% Linebreaks \\ can be used within to get better formatting as desired.
% Do not put math or special symbols in the title.
\title{MDE: Modality Discrimination Enhancement for Multi-modal Recommendation}

% author names and affiliations
% use a multiple column layout for up to three different
% affiliations
\author{\IEEEauthorblockN{Hang Zhou}
\IEEEauthorblockA{
Zhejiang University\\China}
\and
\IEEEauthorblockN{Yucheng Wang}
\IEEEauthorblockA{Nanyang Technological University\\Singapore}
\and
\IEEEauthorblockN{Huijing Zhan}
\IEEEauthorblockA{Singapore University of \\Social Sciences, Singapore}
}

% conference papers do not typically use \thanks and this command
% is locked out in conference mode. If really needed, such as for
% the acknowledgment of grants, issue a \IEEEoverridecommandlockouts
% after \documentclass

% for over three affiliations, or if they all won't fit within the width
% of the page, use this alternative format:
% 
%\author{\IEEEauthorblockN{Michael Shell\IEEEauthorrefmark{1},
%Homer Simpson\IEEEauthorrefmark{2},
%James Kirk\IEEEauthorrefmark{3}, 
%Montgomery Scott\IEEEauthorrefmark{3} and
%Eldon Tyrell\IEEEauthorrefmark{4}}
%\IEEEauthorblockA{\IEEEauthorrefmark{1}School of Electrical and Computer Engineering\\
%Georgia Institute of Technology,
%Atlanta, Georgia 30332--0250\\ Email: see http://www.michaelshell.org/contact.html}
%\IEEEauthorblockA{\IEEEauthorrefmark{2}Twentieth Century Fox, Springfield, USA\\
%Email: homer@thesimpsons.com}
%\IEEEauthorblockA{\IEEEauthorrefmark{3}Starfleet Academy, San Francisco, California 96678-2391\\
%Telephone: (800) 555--1212, Fax: (888) 555--1212}
%\IEEEauthorblockA{\IEEEauthorrefmark{4}Tyrell Inc., 123 Replicant Street, Los Angeles, California 90210--4321}}

% use for special paper notices
%\IEEEspecialpapernotice{(Invited Paper)}

% make the title area
\maketitle

% As a general rule, do not put math, special symbols or citations
% in the abstract
\begin{abstract}
Multi-modal recommendation systems aim to enhance performance by integrating an item's content features across various modalities with user behavior data. Effective utilization of features from different modalities requires addressing two challenges: preserving semantic commonality across modalities (modality-shared) and capturing unique characteristics for each modality (modality-specific). Most existing approaches focus on aligning feature spaces across modalities, which helps represent modality-shared features. However, modality-specific distinctions are often neglected, especially when there are significant semantic variations between modalities. To address this, we propose a \textbf{M}odality \textbf{D}istinctiveness \textbf{E}nhancement (MDE) framework that prioritizes extracting modality-specific information to improve recommendation accuracy while maintaining shared features. MDE enhances differences across modalities through a novel multi-modal fusion module and introduces a node-level trade-off mechanism to balance cross-modal alignment and differentiation. Extensive experiments on three public datasets show that our approach significantly outperforms other state-of-the-art methods, demonstrating the effectiveness of jointly considering modality-shared and modality-specific features.\end{abstract}

% no keywords

% For peer review papers, you can put extra information on the cover
% page as needed:
% \ifCLASSOPTIONpeerreview
% \begin{center} \bfseries EDICS Category: 3-BBND \end{center}
% \fi
%
% For peerreview papers, this IEEEtran command inserts a page break and
% creates the second title. It will be ignored for other modes.
\IEEEpeerreviewmaketitle

\section{Introduction}
With the surge of massive multimedia information \cite{arumugam2022multimodal,xu2024audio,li2023revisiting} across various online platforms, e.g., e-commerce applications and content-sharing communities, multi-modal recommendation systems \cite{zhan20213,zhao2023hierarchical,recsurvey1} have gained increasing popularity. By leveraging the combined information from multiple modalities along with users' historical behavior patterns, the system can effectively capture user preferences for better recommendation. To enhance the quality of recommendations, the key objective is to design an effective approach for integrating comprehensive and complementary information from multiple modalities.

To fully leverage the multi-modality information, a key challenge persists: how to strike a good trade-off between both the semantic commonalities (modality-shared) and the unique characteristics (modality-specific) of each modality. Currently, most existing works \cite{mgcn, lgmrec} have mainly focused on the modality-shared knowledge, which is often achieved through cross-modality alignment using contrastive learning \cite{jaiswal2020survey} and using data augmentation techniques to generate multiple views of the same node, maximizing agreement between these views \cite{mmgcl, slmrec}.

Although these approaches have shown the effectiveness of leveraging the multiple modalities, modality-specific information, which is critical for capturing the unique characteristics of items, is frequently overlooked. For instance, image features of a product typically describe its shape and color, while text features convey its function and price. These distinct types of information can provide valuable recommendations from different perspectives. Although recent research \cite{drepmrec, lattice} has acknowledged the importance of preserving the modality-specific semantics during cross-modal alignment, they still face a common limitation: the inability to fully exploit and further enhance the unique and discriminative information (i.e., modality discrimination) of each modality.

\begin{figure*}[t]
\centering
\includegraphics[width=0.85\textwidth]{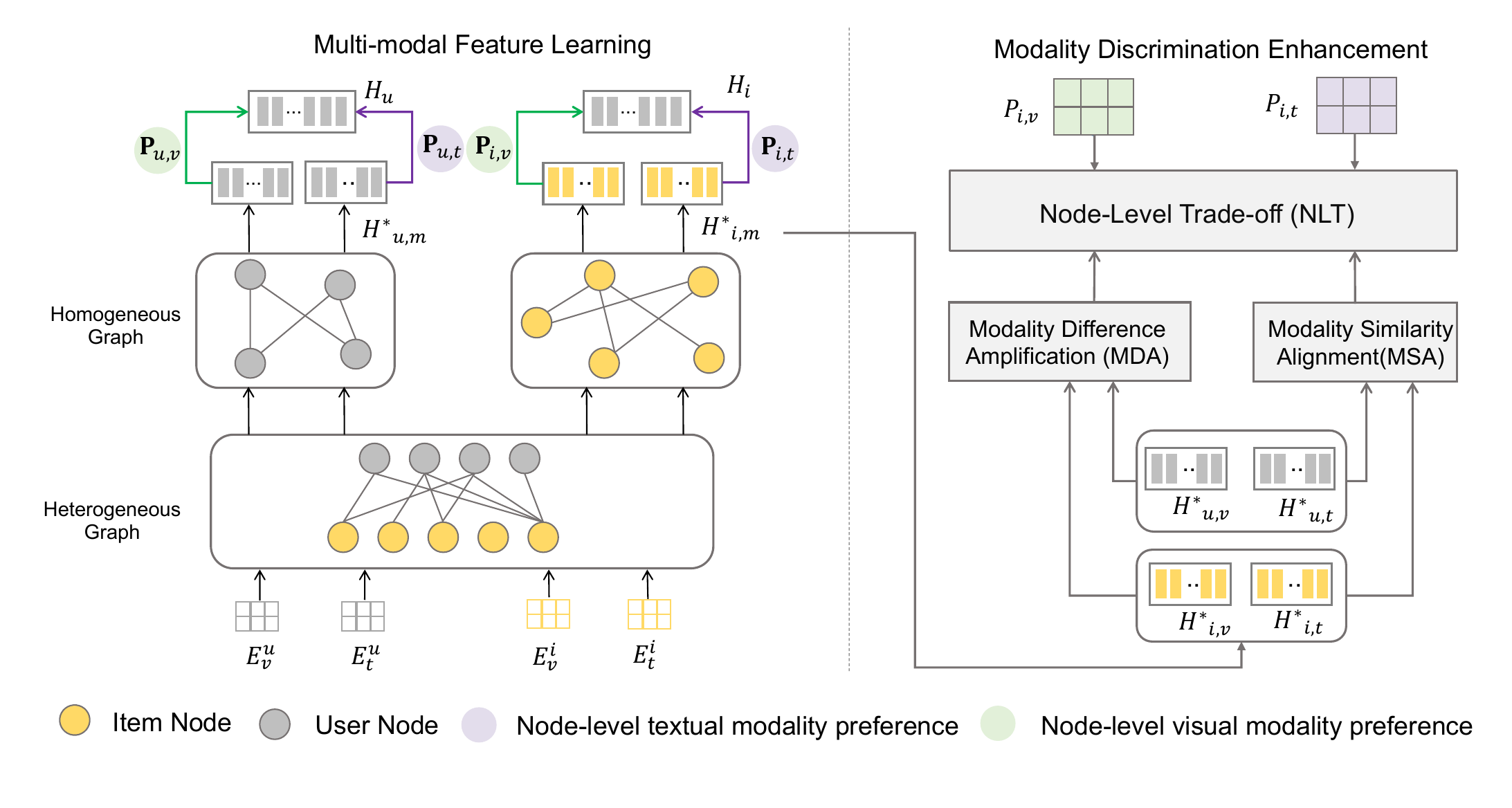} % Reduce the figure size so that it is slightly narrower than the column.
\caption{The framework of the proposed MDE.}
\label{framework}
\end{figure*}

To leverage both modality-shared and modality-specific information for effective recommendation, we propose a novel framework, Modality Discrimination Enhancement (MDE). This framework aims to extract and enhance modality-specific information by amplifying differences between modalities while simultaneously maintaining modality-shared information through modality alignment. Notably, modality alignment inherently conflicts with the amplification of modality differences. To resolve this, we introduce a node-level trade-off mechanism based on learnable modality preferences to balance these two objectives. Specifically, as illustrated in Fig.~\ref{framework}, we first learn multi-modal feature representations by constructing a heterogeneous user-item graph, along with homogeneous graphs for the user-user co-occurrence and item-item semantic similarity. We then introduce a node-level learnable preference weight to effectively fuse the multi-modal features. Finally, we enhance modality discrimination and balance modality-shared and modality-specific information by introducing weighted modality discrimination and modality alignment losses, with the learned modality preferences determining the trade-off. We hypothesize that for a given node, if its preferences across different modalities show larger variations, we should prioritize enhancing modality discrimination, and vice versa. Extensive experiments conducted on several large-scale benchmarks demonstrate the necessity and effectiveness of our modality discrimination method.

\section{Methodology}
\subsection{Problem Definition}
Assume we have a set of users $\mathcal{U}=\left\{u_{i}\right\}_{i=1}^{M}$ and a set of items $\mathcal{I}=\left\{i_{t}\right\}_{t=1}^{N}$. For each item, it has different types of multimedia features (i.e., generated from pre-trained models), denoted as $e_m^i$, where $m\in \mathcal{M}$, where $\mathcal{M}$ refers to the visual and textual modalities, respectively. We then denote the modality feature matrix for items as $\mathbf{E}_m^{i} \in \mathbf{R}^{|\mathcal{I}| \times d_{m}}$ under modality $m$, where $d_m$ denotes the feature dimension. The user historical interaction matrix is denoted as $\mathcal{R} \in \{0,1\}^{|\mathcal{U}| \times|\mathcal{I}|}$,  in which the $r_{u, i}=1$ means the user $u$ has interacted with item $i$, we construct a user-item interaction graph $\mathcal{G}=\{\mathcal{U} \cup \mathcal{I}, \mathcal{E}\}$, where $\mathcal{E}=\left\{(u, i) \mid u \in \mathcal{U}, i \in \mathcal{I}, R_{u i}=1\right\}$ denotes the set of edges. The task of multi-modal recommendation is to learn a prediction function to forecast the preference score $\hat{y}_{u, i}$, which indicates the user $u$'s preference towards the item $i$. Formally, 
\begin{equation}
    \hat{y}_{u, i}=f_{\theta}\left(u, i, \mathcal{G}, \{\mathbf{E}_m^{i}\}_{m \in \mathcal{M}}\right),
\end{equation} 
where $\theta$ refers to model parameters.

\subsection{Multi-modal Feature Learning}
\noindent \textbf{Heterogeneous Graph construction.}
Graph neural networks have been used in this work due to their effectiveness in recommend system to construct user-item and modality-aware auxiliary graphs \cite{hamilton2017inductive, kipf2016semi}. Inspired by \cite{dragon}, we construct both heterogeneous (user-item graph) and homogeneous graphs (user-user graph and item-item graph) to learn high-order connectivity between users and items. Let $\mathbf{E}_{m}^{u}$ and $\mathbf{E}_{m}^{i}$ denote the modal feature matrices for users and items, respectively. These matrices are initialized randomly for users and using features extracted from pre-trained models for items. Here $m\in \{v,t\}$ with $v$ refers to the visual modality and $t$ refers to the textual modality. We adopt LightGCN \cite{lightgcn} to perform message propagation and the forward propagation at the $l+1$ graph convolution layer is formulated as follows:
\begin{equation}
    \mathbf{E}_{m}^{l+1}=\left(\mathbf{D}^{-\frac{1}{2}} \mathbf{A} \mathbf{D}^{-\frac{1}{2}}\right) \mathbf{E}_{m}^{l},
\end{equation}
here $\mathbf{A}$ is the adjacency matrix of the user-item interaction graph $\mathcal{G}$ and $\mathbf{D}$ is the diagonal degree matrix of $\mathbf{A}$ with $\mathbf{D}_{i i}=\sum_{j} \mathbf{A}_{i j}$. Here $\mathbf{E}_m^0$ is initialized as the concatenation of $\mathbf{E}_m^u$ and $\mathbf{E}_m^i$, denoted as 
$\mathbf{E}_{m}^{0}=\mathbf{E}_{m}^{u}||\mathbf{E}_{m}^{i}$. The final feature representation from the heterogeneous graph $\overline{\mathbf{H}}_{m}$ is the average of the embedding of all layers:
\begin{equation}
    \overline{\mathbf{H}}_{m}=\frac{1}{L+1}\sum_{l=0}^{L} \mathbf{E}_{m}^{l}.
    \label{eqhg}
\end{equation}
\noindent \textbf{Homogeneous Graph construction.}
To better explore higher-order relationships between users and items, we also construct homogeneous graphs for both users and items. Drawing the inspiration from \cite{dualgnn,lattice}, we develop a user-user co-occurrence graph and an item-item similarity graph. The user-user co-occurrence graph is created by calculating the number of commonly interacted items between users. For each user, we sample the top-$K$ frequent co-occurred users, resulting in the generation of the graph $\mathcal{G}^u=\{\mathcal{G}_v^u,\mathcal{G}_t^u\}$. In contrast, the item-item similarity graph is constructed by computing the cosine similarity between item features, from which the top-$K$ similar items are sampled to create the item-item similarity graphs, $\mathcal{G}^i = \{\mathcal{G}_{v}^i,\mathcal{G}_{t}^i\}$. We then apply a simple graph convolution to these homogeneous graphs:
\begin{equation}
    \mathbf{H}^*_m=\mathbf{A}_m \overline{\mathbf{H}}_{m},
\end{equation}
where $\mathbf{A}_m=\{\mathbf{A}_m^u, \mathbf{A}_m^i\}$ is the adjacency matrix of the homogeneous graph $\mathcal{G}^u$ and $\mathcal{G}^i$. It is important to note that the graph convolution operation is applied to each modality of the homogeneous graph using its respective adjacency matrix.
%Note that $\mathcal{G}_{u,v}=\mathcal{G}_{u,t}=\mathcal{G}_{u}$. $\overline{H}_{n,m}$ is the corresponding part obtained by separating from $\overline{H}_{m}$ (Equation \ref{eqhg}) according to node type and modality type.

\noindent \textbf{Multi-modal Feature Fusion and Prediction.}
The significance of different modalities often varies in the context of recommendations. For user nodes, individual users may exhibit distinct modality preferences, while for item nodes, one modality may offer more informative insights than another. To more accurately represent these diverse forms of modality preference, we fuse the multi-modal features ${\mathbf{H}}^*_{i,v}$, ${\mathbf{H}}^*_{i,t}$ using a learnable node-level preference. Here, we take the multi-modal feature fusion for the item feature matrix $\mathbf{H}_{i,v}*$ and $\mathbf{H}_{i,t}^*$ as an example. The fusion process for the user feature matrix is denoted as ${\mathbf{H}}_{u}$, which is conducted in a similar manner.
\begin{equation}
\label{eq:fusion}
\begin{aligned}
    {\mathbf{P}}_{i,v} &=\text{softmax}({\mathbf{W}}_{i})[:,0],\\
    {\mathbf{P}}_{i,t} &= \text{softmax}({\mathbf{W}}_{i})[:,1],\\
    {\mathbf{H}}_{i}&={\mathbf{P}}_{i,v} {\mathbf{H}}^*_{i,v} \| {\mathbf{P}}_{i,t}{\mathbf{H}}^*_{i,t},
\end{aligned}
\end{equation}
where ${\mathbf{W}}_{i} \in \mathbf{R}^{|\mathcal{I}| \times 2}$ is a learned parameter matrix. 

After the multi-modal feature fusion for both the users and items, the preference score $\hat{y}_{u, i}$ towards the user $u$ towards the item $i$ is calculated as $\hat{y}_{u, i} = \mathbf{h}_{u}^T \cdot \mathbf{h}_{i}$, where $\mathbf{h}_{u}$ and $\mathbf{h}_{i}$ are rows in ${\mathbf{H}}_{u}$ and ${\mathbf{H}}_{i}$, respectively. The items are ranked in descending order based on the computed preference scores to generate the final recommendation list.

\subsection{Modality Discrimination Enhancement}
\noindent \textbf{Modality Difference Amplification (MDA).}
To extract the modality-specific information, we enhance the discrimination between modalities. Specifically, we maximize the disparity between modality features before the final fusion step:

\begin{equation}
\begin{aligned}
    \mathbf{H}_{diff} &= -|\mathbf{H}^*_{i,v}-\mathbf{H}^*_{i,t}|\odot \mathbf{W}_{diff},\\
    \mathcal{L}_{diff} &= \sigma_{diff}\cdot ||\mathbf{H}_{diff}||^2,
\end{aligned}
\end{equation}
where $|\cdot|$ denotes the absolute value function and $||\cdot||$ represents the L2 norm. $\odot$ denotes the Hadamard product. $\sigma_{diff}$ is a hyper-parameter and $\mathbf{W}_{diff}$ is the node-level weight matrix, which will be introduced in the following section. By minimizing $\mathcal{L}_{diff}$, we can maximize the differences between modality features, thereby preserving the distinctive and informative information in each modality.

\noindent \textbf{Modality Similarity  Alignment (MSA).}
Although we prioritize extracting modality-specific information by maximizing modality differences, how to retain modality-shared information remains essential. We achieve this by the modality similarity alignment, which maps the features of different modalities into a common feature space. We leverage contrastive learning \cite{jaiswal2020survey} to perform the modality alignment:
\begin{equation}
\label{eq:2}
\begin{aligned}
    L_{cl}^{v \to t} &= \sum_{i \in \mathcal{I}}-\log \frac{\exp \left(h_{v}^{i} \cdot h_{t}^{i} / \tau\right) \odot \mathbf{W}_{cl}}{\sum_{i^{\prime} \in \mathcal{I}} \exp \left(h_{v}^{i} \cdot h_{t}^{i^{\prime}} / \tau\right)},\\
    L_{cl}^{t \to v} &= \sum_{i \in \mathcal{I}}-\log \frac{\exp \left(h_{t}^{i} \cdot h_{v}^{i} / \tau\right) \odot \mathbf{W}_{cl}}{\sum_{i^{\prime} \in \mathcal{I}} \exp \left(h_{t}^{i} \cdot h_{v}^{i^{\prime}} / \tau\right)},\\
    L_{cl} &= \sigma_{cl} \cdot (L_{cl}^{t \to v} + L_{cl}^{t \to v}),
    \end{aligned}
\end{equation}
where $\mathbf{W}_{cl}$ is the node-level weight matrix and $\sigma_{cl}$ is the hyper-parameter.

\noindent \textbf{Node-Level Trade-off (NLT).}
In this section, we discuss the trade-off component to balance the modality difference amplification and the modality alignment with these two weight matrices, $\mathbf{W}_{diff}$ and $\mathbf{W}_{cl}$, defined as follows:
\begin{equation}
\begin{aligned}
    \mathbf{W}_{diff} &= f_b(|{\mathbf{P}}_{i,v}-{\mathbf{P}}_{i,t}|),\\
    \mathbf{W}_{cl} &= f_b(1 - |{\mathbf{P}}_{i,v}-{\mathbf{P}}_{i,t}|),
    \end{aligned}
\end{equation}
where $f_b(\cdot): \mathbf{R}^{|\mathcal{I}| \times 1} \to \mathbf{R}^{|\mathcal{I}| \times d}$ is a broadcast function and $d$ is the dimension of features. We leverage the difference of modality preference weight, as shown in Eq.~\ref{eq:fusion}, to determine the trade-off weight. The assumption here is that a larger difference in modality preference implies more importance for modality difference amplification. Conversely, a smaller difference in modality preference indicates a higher weight for the modality alignment, as it suggests that different modalities contribute similarly to the recommendation for that node.  %As shown in equation \ref{eq:1}, \ref{eq:2} and \ref{eq:3}, if a node has a strong modality preference, that node will tend to increase the modality difference more. Otherwise the node will tend to do more modality alignment. 

\subsection{Optimization}
We utilize the Bayesian Personalized Ranking (BPR) Loss \cite{rendle2012bpr} as the optimization objective:
 \begin{equation}
     \mathcal{L}_{{g}}=\sum_{\left(u, i, i^{\prime}\right) \in \mathcal{B}}-\log \left(\sigma\left(h_{u}^T \cdot h_{i}-h_{u}^T \cdot h_{i^{\prime}}\right)\right),
 \end{equation}
where each triple $\left(u, i, i^{\prime}\right)$ satisfies $R_{u i}=1$ and $R_{u i^{\prime}}=0$. $\sigma$ is the sigmoid function. We also introduce two additional modality-specific loss $\mathcal{L}_{{v}}$ and $\mathcal{L}_{{t}}$:
\begin{equation}
     \mathcal{L}_{{v}}=\sum_{\left(u, i, i^{\prime}\right) \in \mathcal{B}}-\log \left(\sigma\left(h^*_{u,v} \cdot h^*_{i,v}-h^*_{u,v} \cdot h^*_{i^{\prime},v}\right)\right),
 \end{equation}
Note that $\mathcal{L}_{{t}}$ is omitted due to space constraints, as it is calculated in a similar manner. The final objective function to be optimized is shown as below:
\begin{equation}
     \mathcal{L}=\mathcal{L}_{g} + \mathcal{L}_{v} + \mathcal{L}_{t} + \mathcal{L}_{diff} + \mathcal{L}_{cl} + \sigma_{reg}\|{\Theta}\|^{2},
\end{equation}
where ${\Theta}$ represents the model parameters and $\sigma_{reg}$ is the regularization coefficient.

\begin{table*}[]
\vspace{-2em}
\caption{The overall performances on several benchmarks.}
\resizebox{\textwidth}{24mm}{\begin{tabular}{l|llll|llll|llll}
\hline
\textbf{Datasets}  & \multicolumn{4}{c}{Baby}          & \multicolumn{4}{c}{Sports}        & \multicolumn{4}{c}{Clothing}      \\
\hline
\textbf{Metrics}   & REC   & PREC   & MAP   &  NDCG   & REC   & PREC   & MAP   &  NDCG & REC   & PREC   & MAP   &  NDCG   \\
\hline
\hline
VBPR    & 0.0265                & 0.0059                & 0.0134                & 0.0170                & 0.0353                & 0.0079                & 0.0189                & 0.0235                & 0.0186 & 0.0039 & 0.0103 & 0.0124 \\
MMGCN   & 0.0240                & 0.0053                & 0.0130                & 0.0160                & 0.0216                & 0.0049                & 0.0114                & 0.0143                & 0.0130 & 0.0028 & 0.0073 & 0.0088 \\
GRCN    & 0.0336                & 0.0074                & 0.0182                & 0.0225                & 0.0360                & 0.0080                & 0.0196                & 0.0241                & 0.0269 & 0.0056 & 0.0140 & 0.0173 \\
DualGNN & 0.0322                & 0.0071                & 0.0175                & 0.0216                & 0.0374                & 0.0084                & 0.0206                & 0.0253                & 0.0277 & 0.0058 & 0.0153 & 0.0185 \\
SLMRec  & 0.0343                & 0.0075                & 0.0182                & 0.0226                & 0.0429                & 0.0095                & 0.0233                & 0.0288                & 0.0292 & 0.0061 & 0.0163 & 0.0196 \\
BM3     & 0.0327                & 0.0072                & 0.0174                & 0.0216                & 0.0353                & 0.0078                & 0.0194                & 0.0238                & 0.0246 & 0.0051 & 0.0135 & 0.0164 \\
FREEDOM & 0.0374                & {\underline{0.0083}} & 0.0194                & 0.0243                & 0.0446                & 0.0098                & 0.0232                & 0.0291                & 0.0388 & 0.0080 & 0.0211 & 0.0257 \\
DRAGON  & {\underline{0.0374}} & 0.0082                & {\underline{0.0202}} & {\underline{0.0249}} & {\underline{0.0449}} & {\underline{0.0098}} & {\underline{0.0239}} & {\underline{0.0296}} & \underline{0.0401} & \underline{0.0083} & \textbf{0.0225} & \textbf{0.0270} \\
\hline
\textbf{MDE} &  \textbf{0.0414}      & \textbf{0.0091}  &  \textbf{0.0219}  & \textbf{0.0273} & \textbf{0.0479}  & \textbf{0.0105}  & \textbf{0.0256} & \textbf{0.0317} & \textbf{0.0402} & \textbf{0.0084} & \underline{0.0219} & \underline{0.0267}\\
\hline
\end{tabular}}

\label{overall performance}
\end{table*}

\section{Experiment}

\subsection{Datasets and Evaluation Setting}
We evaluate our proposed model using three widely recognized Amazon datasets: Baby, Sports, and Clothing \cite{datasets}. We follow the experimental settings established in previous work \cite{datasets2}. The proposed model and all baseline models are implemented using MMRec \cite{survey}. For all models, we fix the embedding size for both users and items at 64, set the training batch size to 2048, use Adam \cite{adam} as the optimizer, and apply the Xavier method \cite{xavier} to initialize the model parameters. The hyperparameters $\sigma_{cl}$, $\sigma_{diff}$ and $\sigma_{reg}$ are determined through grid search on the validation set. We set the number of GCN layers $L$ in the heterogeneous graph as 2. The values of $K$ for constructing the item similarity graph and the user co-occurrence graph are set to 10 and 40, respectively.
 
\subsection{Performance Comparison}
Table \ref{overall performance} presents the Top-5 recommendation performance achieved by various methods. It can be seen that MDE demonstrates significant improvements over state-of-the-art on both the Baby and Sports datasets. The relatively lower performance on the Clothing dataset can be partly attributed to the instability of user modality preferences in the clothing recommendation scenario, which reduces the effectiveness of the modality-preference-based node-level trade-off employed in MDE.

\subsection{Ablation Study}
\begin{table}[]
\centering
\caption{Comparison of MDE with its variants (w/o key components).}
\begin{tabular}{l|l|l|l}
\hline
\hline
                                  & \textbf{Baby} & \textbf{Sports} & \textbf{Clothing} \\
\hline
\textbf{Components}                        & \textbf{R@5}              & \textbf{R@5}              & \textbf{R@5}                \\
\hline                         
%w/o extra BPR                & 0.0387          & 0.0413            & 0.0337              \\     
w/o MDA          & 0.0392          & 0.0444           & 0.0386               \\
w/o MSA            & 0.0394          & 0.0454            & 0.0394               \\
w/o NLT & 0.0385        & 0.0443        & 0.0373           \\
\hline
\textbf{MDE}       & \textbf{0.0414}    & \textbf{0.0479}    & \textbf{0.0402} \\
\hline
\hline
\end{tabular}

\label{ablation}
\end{table}
We conduct ablation experiments to investigate the impact of each individual module on overall model performance. Specifically, we evaluate the performance without the Modality Difference Amplification (MDA), Modality Similarity Alignment (MSA), and Node-Level Trade-Off (NLT) components to assess the framework's effectiveness in the absence of specific modules. The results reported in Table \ref{ablation} reveal several insights based on Recall@5 (R@5)": (1) The performance of the variant ``w/o MDA'' demonstrates that enhancing modality differences is crucial for effectively leveraging multi-modal features. (2) The ``w/o MSA'' results indicate that neglecting modality similarity alignment without constraints leads to a significant drop in performance. (3) The ``w/o NLT'' results confirm the advantages of the proposed adaptive node-level trade-off strategy.
\begin{figure}[htb]
\centering
\includegraphics[width = 0.5\textwidth]{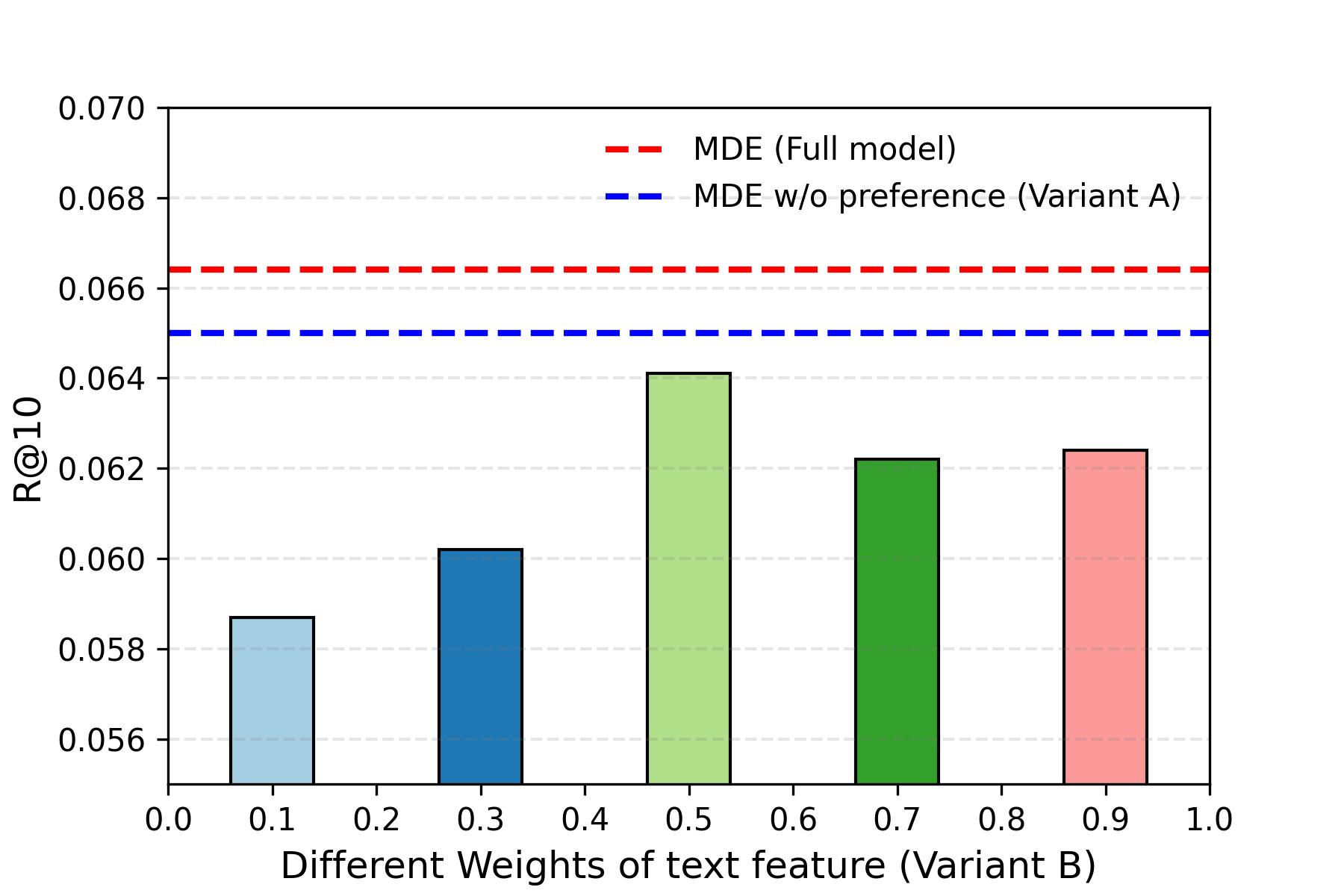} 
\vspace{-2pt}
\caption{Performance of different trade-off mechanism on Baby dataset}
\label{bar} 
\end{figure}
\subsection{Trade-off Mechanism Analysis}
To further assess the effectiveness of the proposed preference-based trade-off mechanism, we compare MDE with the following two variants on Baby dataset: (A) We fix the same modality preference for all nodes while maintaining the preference-based trade-off strategy. In this scenario, the trade-off weights in the loss function are determined by a static modality preference weight setting. The results for variant A are displayed as bar charts in Fig.~\ref{bar}, corresponding to different preference weight settings, where the preference weight for the text modality $w_t$ is fixed and the weight for the visual modality is $1-w_t$. (B) We use node-level independent learnable weights instead of preference-based weights as the trade-off weights in the loss function, with the results for variant B shown as a blue dotted line. From the results in Fig.~\ref{bar}, we observe the following: (1) The contributions of different modalities are uneven, even when individual modality preferences are not considered; on the Baby dataset, the text modality plays a more significant role than images. (2) The results of variants A and B indicate that modality preferences vary individually, suggesting that accounting for node-level modality preferences can enhance performance. (3) The comparison between variant B and MDE demonstrates that the modality-preference-based trade-off mechanism effectively balances modality differentiation and modality alignment.

\section{Conclusion}
In this paper, we propose a novel framework for multi-modal recommendation, named MDE. The proposed approach enhances the modality distinctiveness to effectively extract modality-specific information while maintaining shared features. To achieve this, we introduce the modality difference amplification and modality similarity alignment, aiming to capture both the common and unique knowledge across modalities. To balance these two aspects, we introduce a novel node-level trade-off method based on learnable modality preferences. Extensive experiments on three datasets demonstrate the superiority of our model over state-of-the-art approaches.

% conference papers do not normally have an appendix

% trigger a \newpage just before the given reference
% number - used to balance the columns on the last page
% adjust value as needed - may need to be readjusted if
% the document is modified later
%\IEEEtriggeratref{8}
% The "triggered" command can be changed if desired:
%\IEEEtriggercmd{\enlargethispage{-5in}}

% references section

% can use a bibliography generated by BibTeX as a .bbl file
% BibTeX documentation can be easily obtained at:
% http://mirror.ctan.org/biblio/bibtex/contrib/doc/
% The IEEEtran BibTeX style support page is at:
% http://www.michaelshell.org/tex/ieeetran/bibtex/
\bibliographystyle{IEEEtran}
% argument is your BibTeX string definitions and bibliography database(s)
% \bibliography{bibtex/IEEEexample}
%
% <OR> manually copy in the resultant .bbl file
% set second argument of \begin to the number of references
% (used to reserve space for the reference number labels box)
% \begin{thebibliography}{1}

% \bibitem{IEEEhowto:kopka}
% H.~Kopka and P.~W. Daly, \emph{A Guide to \LaTeX}, 3rd~ed.\hskip 1em plus
%   0.5em minus 0.4em\relax Harlow, England: Addison-Wesley, 1999.

% \end{thebibliography}

% that's all folks
\end{document}